# Negative refraction at deep-ultraviolet frequency in monocrystalline graphite

Jingbo Sun, Ji Zhou*, Lei Kang, Rui Wang, Xianguo Meng, Bo Li, Feiyu Kang and Longtu Li

State Kay Lab of New Ceramics and Fine Processing, Department of Materials Science and Engineering, Tsinghua University, Beijing 100084, China

**Abstract:** Negative refraction is such a prominent electromagnetic phenomenon that most researchers believe it can only occur in artificially engineered metamaterials. In this article, we report negative refraction for all incident angles for the first time in a naturally existing material. Using ellipsometry measurement of the equifrequency contour in the deep-ultraviolet frequency region (typically 254 nm), obvious negative refraction was demonstrated in monocrystalline graphite for incident angles ranging from 20º to 70º. This negative refraction is attributed to extremely strong anisotropy in the crystal structure of graphite, which gives the crystal indefinite permeability. This result not only explores a new route to identifying natural negative-index materials, but it also holds promise for the development of an ultraviolet hyperlens, which may lead to a breakthrough in nanolithography, the most critical technology necessary for the next generation of electronics.

Since the concept of negative-index materials (NIMs) was first proposed by Veselago[1], it has been realised by various mechanisms such as left-handed materials[2-4], photonic crystals[5,6], and artificially indefinite media[7,8]. A well-established route for constructing NIM structures is based on Veselago's theory of left-handed materials (LHM), simultaneous negative permittivity ($\varepsilon$), and magnetic permeability ($\mu$) with different types of metamaterials[9–12]. Recent experimental[13] and theoretical[14] results have also shown negative refraction phenomena in photonic crystals in regimes of negative group velocity and negative effective index above the first band near the Brillouin zone centre ($\Gamma$), without the requirement of an overlap in negative permittivity and permeability. As an alternative approach to NIMs that does not involve the use of LHMs or photonic crystals, negative refraction has been observed in certain classes of uniaxially anisotropic media, a phenomenon that was originally proposed by Lindell *et al.*[15] The general electromagnetic solutions associated with these indefinite media were further solved by Smith and Schruig[16], who indicated that certain classes of indefinite media have identical refractive properties. Negative refraction has been experimentally demonstrated in metamaterials consisting of multilayered structures[7,19] and arrays of metal rods in a dielectric matrix[8,17,21]. Researchers have also shown that, in the long-wavelength limit, the hyperbolic dispersion relation of a three-dimensional (3D) indefinite medium can be valid even for evanescent modes, allowing for use as a hyperlens[18-21]. However, for the frequency range above the visible region, the fabrication of 3D artificial indefinite media is extremely challenging due to limitations in current fabrication technology.

Natural materials may provide a shortcut for bulk optic NIMs as an alternative to metamaterials. Recent research has shown that negative refraction may occur at the interface between a regular and uniaxial medium[22]. Unfortunately, unlike artificial indefinite media that have strong anisotropy, most existing anisotropic crystals possess an absolutely positive dielectric tensor[23-25], so the incident angles and the negative index are too limited to be amenable to practical devices.

Graphite is a semimetal with a uniaxial-layered crystalline structure. As a natural material, the carbon atoms in monocrystalline graphite are arranged in parallel graphene layers formed by a regular hexagonal network of carbon atoms, each of which possesses one free π-electron. Carbon atoms in layers hold each other by strong covalent bonds, while the layers are weakly held together through Van der Waals interactions. This layered structure allows for extremely strong anisotropy between the directions perpendicular and parallel to the atomic plane. This facilitates the establishment of an indefinite tensor of dielectric constant in particular frequency range, which may fit the requirement for negative refraction for certain electromagnetic waves propagating along the atomic plane for all angles. In the present work, we report full-angle negative refraction in the deep-ultraviolet frequency region in monocrystalline graphite.

To demonstrate negative refraction in graphite, we first needed to obtain the spectra of the dielectric constants in graphite, and then chose a suitable operating frequency range in which the dielectric matrix is indefinite. Subsequently, the mapping of the equifrequency contour (EFC) at a certain orientation is performed and refraction properties of different incident angles are analysed according to the EFC. In our experiment, we used HOPG, a closest material to single crystalline graphite, which is manufactured by the Institute of Metal Research, CAS. The mosaic spread of the sample is restricted to 0.689° of the *c*-axis (symmetry axis, perpendicular to the basal plane). The HOPG material was cut into a 1-mm-wide slab along the *c*-axis. Contacting mode AFM was used to acquire the roughness of the optical surface of the slab, and the RMS roughness was determined to be 6.2 nm.

Measurements were performed using a spectroscopic ellipsometer (J. A. Woollam) to determine the pseudodielectric function in three crystal orientations: optic axis is perpendicular to the sample surface; the optic axis is parallel to the sample surface and the plane of incidence; and the optic axis is parallel to the sample surface and perpendicular to the plane of incidence. The $\varepsilon_o$ and $\varepsilon_e$ spectra derived from the measurement results through computation[26] are shown in Fig. 1. In the UV region below 282 nm, $\varepsilon_{o1}$ is negative whereas $\varepsilon_{e1}$ is positive, which may result in full-angle negative refraction.

By studying a uniaxial material, we only need to measure the EFC of the HOPG sample at two different orientations: (1) an orientation in which the optic axis is parallel to the sample surface and perpendicular to the plane of incidence and (2) an orientation in which the optic axis is in the plane of the sample surface and in the plane of incidence. According to the frequency range analysed above, UV light with a wavelength of 254 nm was chosen as the probe light. The incident angle varied from 20° to 70° with a step size of 1°. The ellipsometry parameter $\rho(\theta_i)=\tan(\Psi)\cdot\exp(i\Delta)$ which displays the reflectance ratio of two polarization was collected. By computing Fresnel reflection coefficients $r_p$ and $r_s$ between a uniaxial media and an isotropic media, the raw data obtained from ellipsometer ($r_p/r_s$ vs $\theta_i$) can be transformed into EFC curve of ($k_x$ vs $k_z$) (see the details in supplementary information). The EFC of the HOPG sample was mapped according to the measurement data of the ellipsometer obtained on HOPG, as shown in Figs. 2b and 2e. The latter represents a hyperbolic curve that supplies an obvious confirmation of negative light refraction.

For uniaxial crystals such as graphite, the dielectric constants are characterised by two independent dielectric functions, including the ordinary dielectric function $\varepsilon_o(\lambda)$, which describes light polarised along the carbon-layer planes, and the extraordinary dielectric function $\varepsilon_e(\lambda)$, which describes light polarised perpendicular to the carbon-layer planes. The dielectric property of graphite is determined by its band structure and can be described by different oscillation models or band transition types. For the electric field polarised in the basal plane, π-electrons that arise from each atom can move freely within the two-dimensional graphene layer, which accounts for a negative $\varepsilon_o$ below approximately 0.9 eV (or 1.4 μm). Thus, $\varepsilon_o$ can be expressed as a Drude model with a plasma frequency of approximately 0.9 eV [27]. For photo energies above 0.9 eV, the interband transition (σ-π) along with the intraband transition contributes to $\varepsilon_o$[28]. The intraband part of the dielectric function is described by the Drude model, whereas the other part can be described using a modified Lorentz model[29]. At the photo energy of around 5 eV (240 nm), the hybrid resonance involving both π-and σ-electrons results in a negative $\varepsilon_o$ with a plasma frequency of about 7 eV [28]. For the electric field polarised perpendicular to the basal plane, the transitions between the π bands are forbidden [30], and $\varepsilon_e$ is always positive. Considering that the negative $\varepsilon_o$ below 0.9 eV has a very large imaginary part[27, 31], which may result in a high dielectric loss, we chose the energy region around 5 eV (248 nm) as the operating band (Fig. 1).

When electromagnetic waves propagate and are refracted in a single uniaxial crystalline graphite sample with complex dielectric functions $\varepsilon_x=\varepsilon_y=\varepsilon_o$ and $\varepsilon_z=\varepsilon_e$ (Fig. 2a), the refraction is different for ordinary light and extraordinary light due to the anisotropy. As illustrated in Fig. 2a, if the crystal is oriented with its optic axis parallel to the sample surface and perpendicular to the plane of incidence, the TM incident light can be treated as ordinary light. In this case, the permittivity of the material is isotropic for the electric-field vector **E.** According to Maxwell's equations, the dispersion relationship for TM waves is given by:

$$\frac{k_x^2}{\varepsilon_e} + \frac{k_y^2}{\varepsilon_e} = \frac{\omega^2}{c^2},$$

where c is the velocity of light in a vacuum. As shown in Fig. 2c, this circular EFC results in a normal refraction whose phase and group velocity are parallel.

If the crystal is oriented as shown in Fig. 2d, the incident light is extraordinary light and is modulated by the anisotropic dielectric constants of the crystal. We can then write the EFC as follows:

$$\frac{k_x^2}{\varepsilon_e} + \frac{k_z^2}{\varepsilon_o} = \frac{\omega^2}{c^2}.$$

In this way, the electric-field vector **E** of the refractive light is usually not parallel to the electric-displacement vector **D**. As a result, the Poynting vector, **S$_r$**, pointing in the direction of the energy flow and the wave vector, **k$_r$**, directed along the wavefront normal are not parallel except for the case when **K** = 0 (Fig. 2f). However, the phases of the waves in both media are moving with the same phase velocity along the interface so that the field is continuous over the interface. This requires that the components of the wave vectors parallel to the interface **K** are equal in both media. Consequently, it is possible for the refracted beam to experience normal refraction with regard to **k** but negative refraction with regard to **S**. A uniaxial anisotropic material with $\varepsilon_e > 0$ and $\varepsilon_o < 0$ will exhibit this behaviour for a transverse magnetic (TM) wave propagating along the x-axis for all incident angles. At this time, the EFC of the uniaxial media is hyperbolic, as shown in Fig. 2e, and the Poynting vector S$_r$ is parallel to the normal of the EFC. Negative light refraction is achieved from this hyperbolic dispersion, even though the phase velocity remains positive [32]. The refraction angles can be obtained from the experimental results for different incident angles approximately, as shown in Fig. 3. One can see that the refraction angle of the Poynting vector is negative over the entire range of incident angles.

The aforementioned experiments of negative refraction were also verified by full wave simulations, assuming that the typical dielectric constants are $\varepsilon_o$ = -0.94 + 2.34i and $\varepsilon_e$ = 2.88 + 2.12i for HOPG at the wavelength of 254 nm, as shown in Fig. 1. Using Ansoft HFSS 11, a commercial software based on the finite element theory (FEM), we numerically simulated the propagation of light for different incident angles in graphite, whose principle axis (c-axis) is along z direction. A TM beam (magnetic field is polarised along the y-axis) at the wavelength of 254 nm is incident on the graphite. Perfect E boundary conditions are applied along the y-axis, and open boundaries are applied along the x- and z-axes. The distribution of the magnetic field is plotted in Fig. 4. The results show notable negative refraction at the interfaces, with refraction angles of 26°, 32°, and 45°, corresponding to the incident angles of 30°, 45°, and 60°, respectively. According to the equifrequency contour in Fig. 3, when the incident angles ($\theta_i$) are 30°, 45°, and 60°, the group refraction angles ($\theta_r$) are 16°, 31°, and 40°, respectively. As a result, the corresponding refraction indices of the HOPG sample are shown to be -1.9, -1.37, and -1.35, respectively. These results are in agreement with those obtained from the EFC (Fig. 3).

Considering that a negative $\varepsilon_o$ exists over a relatively wide range in the deep-ultraviolet region below 330 nm, as shown in Fig. 1, full-angle negative refraction can occur over quite a broad band, even though the scale of the negative index will be dependent on the wavelength. Based on the hyperbolic dispersion relation, monocrystalline graphite can be valid for evanescent modes, which preserve the information contained in the high spatial frequencies and are essential for super-resolution applications[21]. As a homogenous material, it avoids severe wave scattering caused by the inner structure in those artificially engineered material. Additionally, because graphite tends to be intercalated with many molecules, ions, or atomic clusters to form new intercalation compounds, the frequency range and the index value of negative refraction can be tunable.

In conclusion, we experimentally demonstrated full-angle negative refraction in a monocrystal of graphite. This negative refraction is attributed to extremely strong anisotropy between the directions perpendicular and parallel to the atomic plane of the graphite structure. As a natural negative-index medium, it avoids complicated design and fine fabrication techniques. As a potential application, a hyperlens in the deep-ultraviolet range based on graphite should enhance state-of-the-art photolithography and break the bottleneck of reducing the feature size of the latest IC technology. More significantly from a fundamental perspective, this method provides a route to find negative-index materials in nature.

Acknowledgements
This work is supported by the National Science Foundation of China under Grant Nos. 90922025, 50632030, 50921061, and 10774087.

-------------------------------------------------------

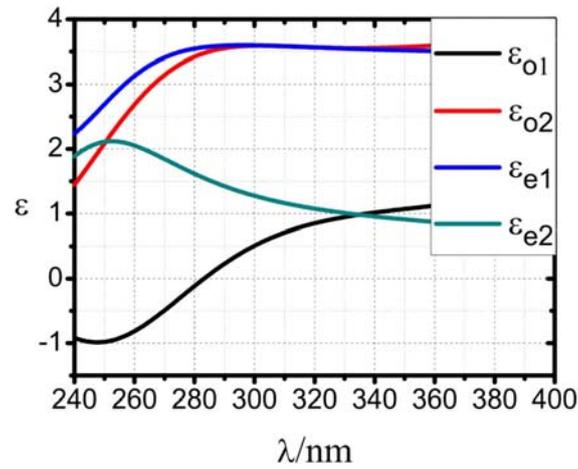

Fig. 1 The spectral dependence of the real ($\varepsilon_{o1}$, $\varepsilon_{e1}$) and imaginary ($\varepsilon_{o2}$, $\varepsilon_{e2}$) parts of the dielectric constants for different orientations in HOPG.

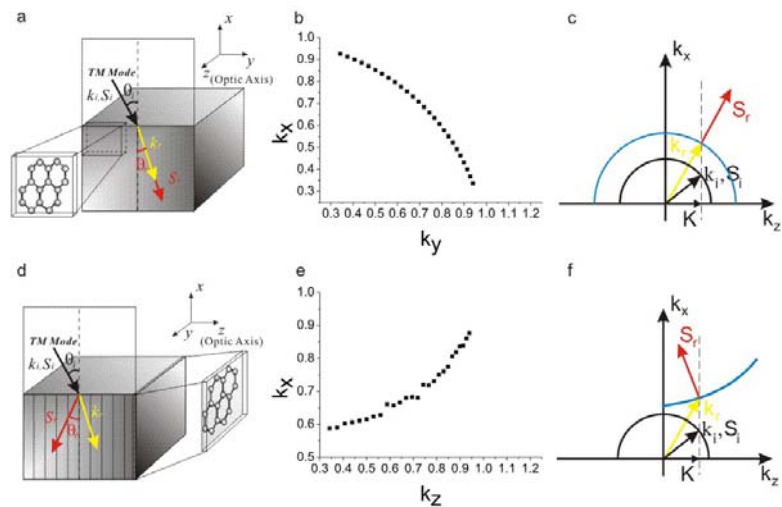

Fig. 2 The orientation of single crystalline graphite and the scheme of refraction as it occurs in graphite. The direction perpendicular to the carbon atomic plane is the optic axis (z-axis). a, The single crystal of graphite is oriented such that the optic axis is parallel to the sample surface and is also perpendicular to the plane of incidence. A TM mode wave is incident on the sample with an incident angle of $\theta_i$ and then refracted at an angle of $\theta_r$. b, The EFC mapped using ellipsometry at this orientation; c, Schematic illustration of the normal refraction of ordinary light at the interface between the free space (circular black EFC) and the uniaxial media (circular blue EFC). The refracted wave vector $k_r$ and Poynting vector $S_r$ can be determined by satisfying Maxwell's theorem. d, The single crystal of graphite is oriented such that the optic axis is in the plane of the sample surface and in the plane of incidence. e, The EFC mapped using ellipsometry at this orientation. f, The negative refraction of extraordinary light at the interface between the free space (circular black EFC) and the uniaxial media (hyperbolic blue EFC).

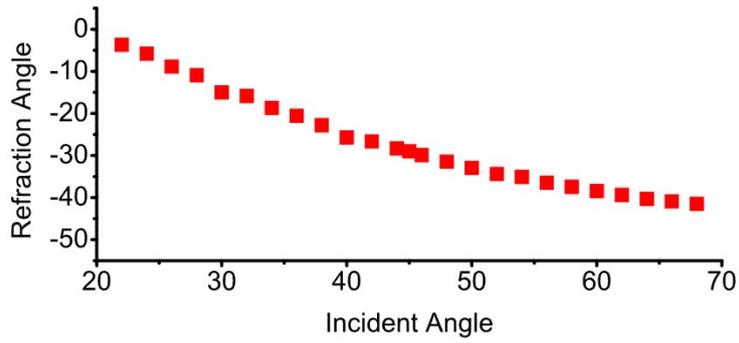

Fig. 3 Refracted angles θ_r for the Poynting vector at various incident angles. The refraction angle can be obtained by computing the normal of the hyperbola.

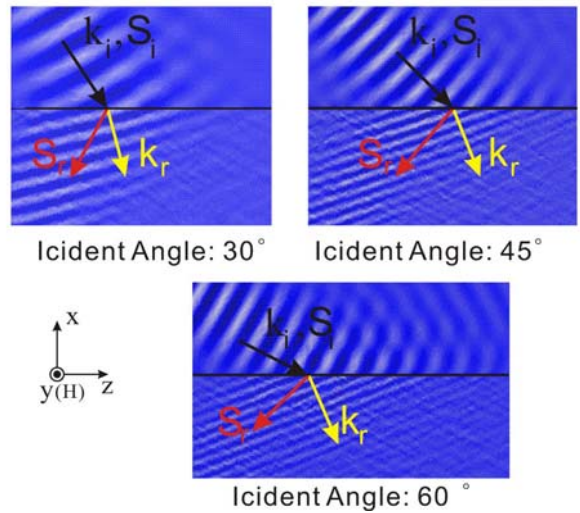

Fig. 4 Finite-element simulation results for TM light at incident angles of 30° (left), 45° (right), and 60° (below) using HFSS. The refracted wave vector (yellow arrow) and Poynting vector (red arrow) can be determined by satisfying the Maxwell's theorem. The Poynting vector is negatively refracted, although the phase vector remains positive.

**Supplementary Information**

1. The hyperbolic EFC of the graphite obtained by ellipsometery

Ellipsometry measures the change in polarization state of light reflected from the surface of a sample. The measured values are expressed as Ψ and Δ, as shown in Fig. 1. These values are related to the ratio of Fresnel reflection coefficients $r_p$ and $r_s$ for p- and s- polarized light, respectively.

$$\rho = \frac{r_p}{r_s} = \tan\psi e^{i\Delta}.$$

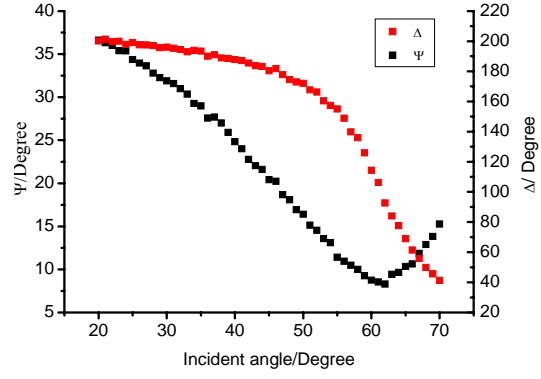

Fig. 1 The average values of five groups raw data measured by ellipsometer

By computing Fresnel reflection coefficients $r_p$ and $r_s$ between a uniaxial media and an isotropic media, the raw data obtained from ellipsometer ($r_p/r_s$ vs $\theta_i$) can be transformed into EFC curve of ($k_x$ vs $k_z$). If the HOPG sample was oriented with its optic axis in the plane of sample surface and incidence (Fig. 1 a), the dispersion relation for TM waves is given by

$$\frac{k_x^2}{\varepsilon_z} + \frac{k_z^2}{\varepsilon_x} = \frac{\omega^2}{c^2}, \qquad (1)$$

where c is the light velocity in vacuum. Following the definition above, we have $\varepsilon_o=\varepsilon_x$, $\varepsilon_e=\varepsilon_z$. The wave vector can be written as

$$k_z = \frac{\sin\theta_i \cdot \omega^2}{c^2}, \qquad (2)$$

$$k_x = \sqrt{\left(\frac{\omega^2}{c^2} - \frac{k_z^2}{\varepsilon_o^2}\right)\varepsilon_e}. \qquad (3)$$

A probe light from an isotropic media incident on the uniaxial sample, whose optic axis is in the plane of sample surface and incidence, makes an angle $\theta_t$ with respect to the surface normal. In this case, the complex reflection coefficients are given by

$$r_p = \frac{\sqrt{\varepsilon_o\varepsilon_e}\cdot\cos(\theta_i) - \sqrt{\varepsilon_o - \sin^2(\theta_i)}}{\sqrt{\varepsilon_o\varepsilon_e}\cdot\cos(\theta_i) + \sqrt{\varepsilon_o - \sin^2(\theta_i)}}, \qquad (4)$$

$$r_s = \frac{\cos(\theta_i) - \sqrt{\varepsilon_o - \sin^2(\theta_i)}}{\cos(\theta_i) + \sqrt{\varepsilon_o - \sin^2(\theta_i)}}. \qquad (5)$$

The raw data obtained from ellipsometer ($r_p/r_s$ vs $\theta_i$) can be transformed into EFC curve of ($k_x$ vs $k_z$) by substituted equations (2) and (3) into equations $r_p/r_s$.

The computation of EFC with normal refraction property is similar to the deduction above except for the use of **reflection coefficients between two isotropic media.**

2. The computation of the refraction angles from the EFC

Assume that a TM light (H is along the y-axis) incident on a uniaxial media with dielectric

constant:

$$\boldsymbol{\varepsilon} = \hat{u}_x \varepsilon_{xx} + \hat{u}_y \varepsilon_{yy} + \hat{u}_z \varepsilon_{zz}.$$

The general format of E field and H field can be written as:

$$\vec{E}(r) = E_0 e^{j(\vec{k} \cdot \vec{r} - wt)}, \quad \vec{H}(r) = H_0 e^{j(\vec{k} \cdot \vec{r} - wt)}.$$

According to Maxwell's equations, we can write:

$$\nabla \times \vec{E} = -\frac{\partial \vec{B}}{\partial t} \qquad \nabla \times \vec{H} = \frac{\partial \vec{D}}{\partial t},$$

$$k \times \vec{E}_0 = \omega \mu_0 \mu_r \vec{H}_0, \tag{6}$$

$$k \times \vec{H}_0 = -\omega \varepsilon_0 \boldsymbol{\varepsilon} \vec{E}_0. \tag{7}$$

Thus, the E field vector is

$$\vec{E}_0 = -\frac{k \times \vec{H}_0}{\omega \varepsilon_0 \boldsymbol{\varepsilon}}$$

$$= -\frac{1}{\omega \varepsilon_0} (k_x \quad 0 \quad k_z) \times (0 \quad H_0 \quad 0) \bullet \begin{pmatrix} \varepsilon_{xx} & 0 & 0 \\ 0 & \varepsilon_{yy} & 0 \\ 0 & 0 & \varepsilon_{zz} \end{pmatrix}^{-1}$$

$$= \frac{1}{\omega \varepsilon_0} \left( \frac{k_z \bullet H_0}{\varepsilon_{xx}} \quad 0 \quad \frac{-k_x \bullet H_0}{\varepsilon_{zz}} \right) \tag{8}$$

Then, we can write down the Poynting vector:

$$\vec{S} = \frac{1}{2} \vec{E} \times \vec{H}^* = \frac{1}{2} \frac{1}{\omega \varepsilon_0} \left( \frac{k_z \bullet H_0}{\varepsilon_{xx}} \quad 0 \quad \frac{-k_x \bullet H_0}{\varepsilon_{zz}} \right) \times (0 \quad H_0 \quad 0)$$

$$= \frac{1}{2} \frac{1}{\omega \varepsilon_0} \left( \frac{k_x \bullet H_0^2}{\varepsilon_{zz}} \quad 0 \quad \frac{k_z \bullet H_0^2}{\varepsilon_{xx}} \right)$$

$$= \frac{1}{2} \frac{1}{\omega \varepsilon_0} \left( \frac{k_x \bullet H_0^2}{\varepsilon_{zz}} \hat{u}_x + \frac{k_z \bullet H_0^2}{\varepsilon_{xx}} \hat{u}_z \right), \tag{9}$$

and its two components along two different directions:

$$S_{rx} = \frac{k_x \bullet H_0^2}{2 \omega \varepsilon_{zz}}, \tag{10}$$

$$S_{rz} = \frac{k_z \bullet H_0^2}{2 \omega \varepsilon_{xx}}.$$

As a result, the refraction angle for the Poynting vector $S_r$ can be determined by

$$\frac{S_x}{S_z} = \frac{\varepsilon_{xx} k_x}{k_z \varepsilon_{zz}}. \qquad (11)$$

By computing the derivative of the EFC equation:

$$\frac{k_x^2}{\varepsilon_z} + \frac{k_z^2}{\varepsilon_x} = \frac{\omega^2}{c^2},$$

we can find

$$-\left(\frac{dk_x}{dk_z}\right)^{-1} = \frac{S_x}{S_z}. \qquad (12)$$

Based on this result, we find that the Poynting vector $S_r$ is parallel to the normal of the EFC, so the refraction angle can be derived by computing the normal of the EFC in Fig. 2. Result shows a part of hyperbola in the ($k_x$, $k_z$) plane in Fig. 2

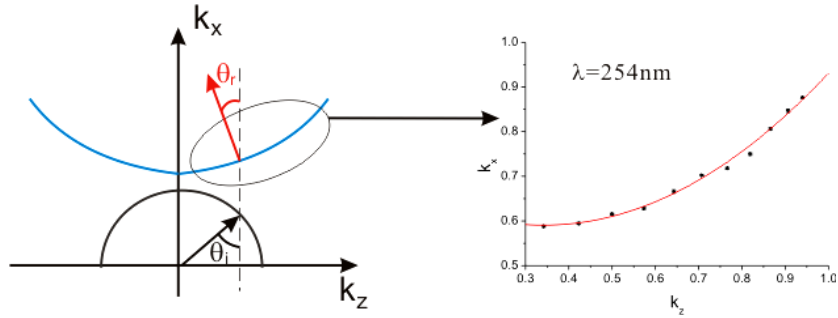

Fig. 2 EFC derived by the raw data and the computation of the refraction angles.